\documentclass[runningheads]{llncs}

\usepackage{graphicx}

\usepackage[utf8]{inputenc}
\usepackage[english]{babel}
\usepackage[table]{xcolor}
\usepackage{array,makecell}
\usepackage{booktabs}
\usepackage{pifont}
\usepackage{tikz}

\usepackage{comment}
\usepackage{todonotes}

\usepackage{hyperref}

\newcommand*\yes{\tikz\draw[black, fill=black](0,0) circle (0.8ex);}
\newcommand*\no{\tikz\draw[black, fill=white](0,0) circle (.8ex);}
\newcommand*\half{%
    \begin{tikzpicture}
        \draw[fill=black] (0,0) circle (0.8ex);
        \draw[fill=white] (0,0)-- (90:0.8ex) arc (90:270:0.8ex) -- cycle ;
\end{tikzpicture}}

\newcommand*\numberoftools{18}

\begin{document}

\title{Continuously Testing Distributed IoT Systems: An Overview of the State of the Art}
\titlerunning{Continuously Testing Distributed IoT Systems}
\author{Jossekin Beilharz\inst{1} \and
Philipp Wiesner\inst{2} \and
Arne Boockmeyer\inst{1} \and 
\\
Lukas Pirl\inst{1} \and 
Dirk Friedenberger\inst{1} \and
Florian Brokhausen\inst{2} \and
\\
Ilja Behnke\inst{2} \and
Andreas Polze\inst{1} \and 
Lauritz Thamsen\inst{3,2}}

\authorrunning{J. Beilharz et al.}

\institute{
Hasso Plattner Institute, University of Potsdam, Germany\\
\email{\{firstname.lastname\}@hpi.de}\\
\email{dirk.friedenberger@guest.hpi.de} \and
Technische Universit{\"a}t Berlin, Germany\\
\email{\{wiesner, florian.brokhausen, i.behnke\}@tu-berlin.de} \and
Humboldt-Universität zu Berlin, Germany\\
\email{lauritz.thamsen@hu-berlin.de}
}

\maketitle

\begin{abstract}
The continuous testing of small changes to systems has proven to be useful and is widely adopted in the development of software systems. 
For this, software is tested in environments that are as close as possible to the production environments. 
When testing IoT systems, this approach is met with unique challenges that stem from the typically large scale of the deployments, heterogeneity of nodes, challenging network characteristics, and tight integration with the environment among others.
IoT test environments present a possible solution to these challenges by emulating the nodes, networks, and possibly domain environments in which IoT applications can be executed.
This paper gives an overview of the state of the art in IoT testing.
We derive desirable characteristics of IoT test environments, compare \numberoftools{} tools that can be used in this respect, and give a research outlook of future trends in this area.

\end{abstract}

\section{Introduction} %
\label{sec:intro}

The Internet of Things (IoT) has the potential to transform our lives by connecting everyday objects to the Internet for smarter cities, factories, houses, and more.
To realize this vision, distributed software systems will need to integrate IoT devices – usually equipped with sensors and actuators – allowing them to continuously monitor and interact with their environments.
These distributed software systems of the IoT will span from devices to clouds and, in many cases, also include intermediate resources at the edge or fog level~\cite{Dastjerdi_2016_FogComputing}.
Examples of distributed IoT systems include those that control and manage traffic and transportation~\cite{Masek_2016_IoTTraffic,Zhao_2021_TrainIoT}, those that enable telemedicine and remote patient monitoring~\cite{Malasinghe_2019_Remote,Gontarska_2021_Telemed}, and those that detect and predict failures as well as optimize processes in urban infrastructures and manufacturing~\cite{Kang_2016_SmartManufacturing,Mohammadi_2018_SmartCityBigML,Geldenhuys_2021_Dependable}.

A major remaining challenge to practically developing and deploying distributed IoT systems is the difficulty of adequately testing them~\cite{Kim_IoTTaaS_2018}.
This is complicated due to a number of factors, including the large number of devices, the heterogeneity of devices, mobile nodes resulting in dynamic topologies, network disconnections and node failures, as well as a tight integration of systems with their respective environments.
At the same time, properly testing IoT systems in application domains such as traffic and transportation management, patient monitoring, and factory processes is absolutely critical.
Consequently, the need for adequate testing of distributed IoT systems has been widely recognized and many solutions have been put forward. 
Prominent examples include hardware testbeds like StarBED~\cite{miyachi_starbed_2006} and FIT-IoT~\cite{adjihFITIoTLABLarge2015}, hybrid approaches such as Chameleon~\cite{keaheyLessonsLearnedChameleon2020}, as well as simulators like IoTSim~\cite{Zeng_IOTSim_2016} and iFogSim \cite{Gupta_iFogSim_2017}.

Hardware testbeds allow to execute actual application code in realistic settings, yet can be limited in terms of scalability and flexibility.
Hybrid test environments address these limitations by incorporating both actual hardware and virtual nodes.
Simulations on the other hand enable to flexibly assess the behavior of distributed applications over various scales and possible infrastructures.
However, they usually lack the ability to evaluate the non-functional properties of actual application code.

All these approaches have in common that it is typically hard to test distributed IoT systems within their actual environment.
Field testing regularly requires a large and coordinated effort, so distributed IoT systems cannot be tested continuously, while lab testing routinely resorts to merely replaying sensor data, so that the distributed IoT systems, despite being equipped to interact with environments, cannot actually influence their domains.
This runs contrary to generally understood and widely adopted principles and best practices of iterative software development, where continuous testing of small changes to systems in environments that mirror production environments as closely as possible is a key mechanism for fast feedback and trust in changes.
We, therefore, argue that there is a significant lack of approaches and tools for continuously testing IoT systems.

In this paper, we compare currently available IoT test environments to provide an overview over the current state of the art and expose the perceived research gap.
For our comparison, we selected test environments that
\begin{enumerate}
  \item focus on testing software systems on geo-distributed, heterogeneous computing infrastructures such as IoT and edge/fog architectures,
  \item allow to run actual system code (i.e., not merely simulating communication),
  \item and have the ability to include virtual nodes, allowing tests at large and various scales (i.e., no hardware-only testbeds).
\end{enumerate}

We only discuss general-purpose test environments of which details have been published (i.e., no proprietary offers such as IoTIFY\footnote{\url{https://iotify.io/}} or AWS IoT Device Simulator\footnote{\url{https://aws.amazon.com/solutions/implementations/iot-device-simulator/}}).

We report the following aspects of IoT test environments in our comparison: how and with which capabilities the tools provide nodes, how the network between nodes is realized, whether domain environments are integrated, as well as general aspects such as project maturity and ongoing development.

The main contributions of this paper are:

\begin{itemize}
    \item A description of key characteristics of IoT test environments, which can be used to distinguish proposed solutions.
    \item A point-by-point comparison of state-of-the-art IoT test environments that meet the outlined selection criteria.
    \item A discussion of current trends and considerable gaps in the state of the art of IoT testing.
    \item An outlook on future work to close these gaps and an overview of our work in this area.
\end{itemize}

The remainder of this paper is structured as follows:
Section~\ref{sec:characteristics} describes central characteristics of test environments.
These are used in Section~\ref{sec:comparison} to evaluate and compare concrete test environments.
Section~\ref{sec:discussion} discusses the results of our comparison.
Section~\ref{sec:outlook} presents the research outlook.
Section~\ref{sec:rw} covers related work.
Lastly, Section~\ref{sec:summary} concludes this paper.

\section{Characteristics of Test Environments} %
\label{sec:characteristics}

Continuously testing IoT systems and applications requires a test environment that reproduces reality as close as possible. 
To classify and compare existing test environments we derive several quantifiable characteristics from generally desirable properties of test environments.

To be able to continuously develop and test distributed IoT systems in iterative software development processes, we need to be able to deploy and run actual code in flexible, yet realistic environments. 
To facilitate large-scale deployments while also allowing the realistic testing of system behavior, we believe the support for both virtualized nodes as well as hardware nodes in test environments is crucial. %
The testing of large-scale deployments further requires the distributability of not only the nodes, but also the network representation and the simulation of the domain environment. 
Another important aspect relating to the three feature dimensions here --- nodes, networks, and the domain environments --- is the meaningful testing of fault tolerance of IoT systems by precisely injecting faults.
Lastly, because IoT systems are inherently integrated tightly with their specific environment through sensors and actuators, we believe that the simulation of the domain environment is a key characteristic for test environments.

The remainder of this section discusses the characteristics that will be used in the following comparison of test environments in Section~\ref{sec:comparison}.
We identified 16 different attributes, which are organized into four overarching categories, regarding general features, the nodes, the network and the domain environment.

\subsection{General}

First, we describe general attributes of test environments. We present the \textit{initial} year of \textit{publication} along with the information if the project is \textit{actively maintained}, which is assessed based on whether there has been a new release or active collaboration (e.g., commits to the repository) in 2021.

As the \textit{maturity} of a project is subjective, we try to formalize it as follows.
An empty circle (\no) denotes the lowest maturity, meaning that the specific test environment only exists as a concept in form of a publication, but no actual tool is available.
We did not assess whether such concepts are actively maintained, as this cannot be sensibly judged.
The second degree of maturity (\half) is reported if the tool is available only as a prototype without good documentation.
If there is a full system available with detailed documentation, we denote it as the third degree of maturity (\yes). 

Next, we classify if a test environment is \textit{offered as a service}.
This indicates whether there is a service where the test environment can be used without manually deploying and operating it.

Lastly, we asses the property \textit{scriptable scenarios}, which is fulfilled if the execution of experiments can be controlled via a script.
With the capability to predefine schemes to alter parameters and characteristics of a simulation at runtime, much more complex scenarios can be implemented.
This is highly important when systematically approaching an investigation with a test environment.

These general information about a test environment can serve as an indicator for the applicability to current challenges, but they are also used to identify recent trends in test environments in Section~\ref{sec:discussion}.

\subsection{Nodes}

An essential aspect of test environments is which type of nodes can be used.
The attributes investigated here determine if a scenario or application of interest can at all be properly implemented or analyzed with a given test environment.

The first attribute, \textit{hardware integration}, classifies the test environments according to their capability to integrate physical hardware nodes.
The availability of hardware integration enables the inclusion of embedded systems and facilitates testing of applications in realistic environments.

The \textit{virtualization type} describes how virtual nodes are represented, namely via virtual machines (V), containerized nodes (C), or a combination of the two (VC).
Depending on the application under test, the differentiation between containerized and virtualized nodes can be crucial. 
Virtual machines enable a more realistic execution environment for the application under test, while containerization is a more light-weight approach.

For the \textit{energy consumption} characteristic, we investigate if a test environment facilitates modeling (or, in the case of hardware nodes, monitoring) the power consumption of nodes and network.
In any use case where energy is a scarce resource, for example for battery-constrained IoT devices, this feature allows testing the effect of software changes to a node's power usage.

The \textit{distributability} describes whether virtual nodes of the test environment can be spread across multiple physical host nodes, enabling large-scale scenarios.

Finally, we investigate the possibility of \textit{fault injection}.
For individual nodes, examples include the purposeful shutdown or internal failure of a given node at a given time.
By simulating such faults, the robustness of an application or network setup towards faults can be tested. 
We analyze the characteristics of distributability and fault injection as well for the network and the domain environment categories.

\subsection{Network}

Regarding network, we first analyze the \textit{network type}, namely how network is emulated within the test environment.
Traffic shaping (TS) allows users to change network parameters, like delay or bandwidth. Examples of this are the Linux Traffic Control (\textit{tc}) or the more advanced NetEm.
Tools based on software-defined networks (SDN) use a virtualized network such as provided by Mininet or MaxiNet.
Lastly, network simulators (NS) can be used to model the underlying network.
In our understanding, network simulators can simulate different kinds of networks, also future ones, without having them physically available. 
Common network simulators are ns-3 or OMNet++ with INET. 

Network \textit{distributability} regards the possibility of the test environment to span the network across multiple physical hosts, leveraging more complex routing schemes in a physical network. For traffic shaping-based approaches this comes naturally if nodes are distributed, for network simulation-based approaches also the simulation has to run in a distributed manner.

\textit{Fault injection} entails active support of the tool to purposefully alter network connections at runtime, e.g., the increase of latency or the loss of packets. Such capabilities are important when comparing fault tolerance of network setups and in general testing of network robustness.

\subsection{Domain Environment}

As IoT applications run in embedded, real-life settings like traffic control, water management or smart homes, simulating the domain environment is of high importance.
First, we asses the general \textit{domain environment support} of a tool, meaning whether there is an API for connecting domain-specific simulators that can interact with the test environment at runtime.
Examples include a traffic simulation, such as SUMO, that can send the coordinates of mobile nodes to the test environment for it to adapt its networking parameterizations.

Similar to the node and network categories, we also asses the \textit{distributability} of the domain environment to see if the execution of the environment can be spread across multiple hosts.

Last, we report the capability of \textit{fault injection} inside the domain environment. 
We define this functionality to be present when the test environment supports to alter the domain environment during runtime in a way that is expected to introduce faults in the application running on the nodes.

\section{Comparison of Test Environments} %
\label{sec:comparison}

Based on the selection criteria defined in Section~\ref{sec:intro}, we selected \numberoftools{} environments for testing IoT applications and evaluated them on the characteristics described in Section~\ref{sec:characteristics}.
Table~\ref{tab:comparison} provides an overview of all evaluated tools.
We clustered the test environments (1) by their ability to integrate real IoT devices in their experiments and (2) by the type of network modeling approach they use.

\newcommand*\rot{\rotatebox{90}}
\newcolumntype{C}[0]{>{\centering}b{0.45cm}}
\newcolumntype{L}[0]{>{\centering}b{0.75cm}}

\begin{table}[t]
    \centering
    \caption{An overview of test environments for IoT systems. A gray background marks works where code is not openly available. All characteristics are described in Section~\ref{sec:characteristics}.}

\begin{tabular}{@{} l|L*{4}C|*{5}C|L*{2}C|*{3}C|c}
& \multicolumn{5}{c|}{} & \multicolumn{5}{c|}{} & \multicolumn{3}{c|}{\tiny{}} & \multicolumn{3}{c|}{Domain} \\
   & \multicolumn{5}{c|}{General} & \multicolumn{5}{c|}{Nodes} & \multicolumn{3}{c|}{Network} & \multicolumn{3}{c|}{Env.} & \\
   
\rowcolor{gray!30} \cellcolor{white}
   & \rot{Initial Publication} & \rot{Actively Maintained} & \rot{Maturity} & \rot{Offered as a Service} 
   & \rot{Scriptable Scenarios} & \rot{Hardware Integration} & \rot{Virtualization Type} 
   & \rot{Energy Consumption} & \rot{Distributability} & \rot{Fault Injection} & \rot{Network Type} & \rot{Distributability} & \rot{Fault Injection} & \rot{Domain Env. Support } & \rot{Distributability} & \rot{Fault Injection} &\cellcolor{white} \\

        \cline{1-17}
        
EMU-IoT~\cite{ramprasadEMUIoT2019}                             & 2019 & \no     & \half & \no & \yes     & \no & C & \no & \yes & \no     & \no & -  & -         & \no & - & - &\\
ELIoT~\cite{makinenELIoTDesignEmulated2017a}                     & 2017 & \no     & \half & \no & \no     & \no & C & \no & \yes & \no     & \no & - & -         & \no & - & - &\\
\rowcolor{gray!10} IOTier~\cite{Nikolaidis_IOTier_2021}                 & 2021 & -     & \no & \no & \yes     & \no & C & \no & \yes & \yes     & TS & \yes & \yes     & \no & - & - &\cellcolor{white}\\
Fogify~\cite{Symeonides_Fogify_2020}                             & 2020 & \yes    & \yes & \no & \yes     & \no & C & \yes & \yes & \yes     & TS & \yes & \yes & \no & - & - &\\
MockFog~\cite{hasenburgMockFogAutomatedExecution2021}                             & 2019 & \yes     & \half & \no & \yes     & \no & V & \no & \yes & \yes     & TS & \yes & \yes     & \no & - & - &\\
Blockade~\cite{blockade}                                                    & 2014 & \yes     & \yes & \no & \no     & \no & C & \no & \no & \yes     & TS & \no & \yes     & \no & - & - &\\
Distem~\cite{sarzyniec2013design}                                 & 2013 & \no     & \yes & \no & \yes     & \no & C & \no & \yes & \yes     & TS & \yes & \yes     & \no & - & - &\\
Fogbed~\cite{coutinhoFogbedRapidPrototypingEmulation2018}                             & 2018 & \no     & \half & \no & \no     & \no & C & \no & \yes & \no     & SDN & \yes & \no     & \no & - & - &\\
EmuFog~\cite{mayerEmuFogExtensibleScalable2017a}                                 & 2017 & \no     & \half & \no & \no     & \no & C & \no & \yes & \yes     & SDN & \yes & \yes     & \no & - & - &\\
Dockemu~\cite{petersenDockemuEvolutionNetwork2019}             & 2015 & \no     & \half & \no & \yes     & \no & C & \no & \yes & \no     & NS & \no & \no     & \no & - & - &\\
EmuEdge~\cite{zeng2019emuedge}                                 & 2019 & \no     & \half & \no & \yes     & \yes & VC & \no & \no & \yes     & TS & \no & \yes     & \no & - & - &\\
Héctor~\cite{behnke2019hector}                                 & 2019 & \no     & \half & \no & \yes     & \yes & V & \no & \no & \yes     & TS & \no & \yes     & \no & - & - &\\
\rowcolor{gray!10} Sendorek et al.~\cite{Sendorek_SDNTestbedForIoT_2018} & 2018 & -     & \no & \no & \yes     & \yes & V & \no & \no & \no     & SDN & \no & \no     & \no & - & - &\cellcolor{white}\\
Chameleon~\cite{keaheyLessonsLearnedChameleon2020}             & 2015 & \yes     & \yes & \yes & \no     & \yes & V & \no & \yes & \no     & SDN & \yes & \no     & \no & - & - &\\
StarBED~\cite{miyachi_starbed_2006}                             & 2002 & \yes     & \yes & \yes & \yes     & \yes & V & \no & \yes & \no     & SDN & \yes & \no     & \no & - & - &\\
\rowcolor{gray!10} UiTiOt~\cite{ly-trongUiTiOtV3Hybrid2018a}                & 2017 & -     & \no & \no & \no     & \yes & C & \no & \yes & \no     & NS & \yes & \yes     & \no & - & - &\cellcolor{white}\\
\rowcolor{gray!10} WHYNET~\cite{zhouWHYNETHybridTestbed2006}     & 2006 & -     & \no & \no & \yes     & \yes & V & \yes & \yes & \no     & NS & \yes & \yes     & \no & - & - &\cellcolor{white}\\
\rowcolor{gray!10} MobiNet~\cite{mahadevanMobiNetScalableEmulation2006}             & 2005 & -     & \no & \no & \no     & \yes & V & \no & \yes & \no     & NS & \no & \no     & \no & - & - &\cellcolor{white}\\
    \end{tabular}

\label{tab:comparison}
\end{table}

\subsection{Test Environments without Hardware Integration}

First, we cover test environments that emulate IoT environments without the possibility to integrate real IoT devices in experiments. These test environments are further categorized based on whether they use network simulation, SDN-based solutions, or simple traffic shaping to emulate realistic network traffic.

\subsubsection{Using no Network Model.}

EMU-IoT~\cite{ramprasadEMUIoT2019} is a container-based test environment with a focus on defining, orchestrating and monitoring reproducible experiments.
Although the authors describe the many challenges faced by developing IoT test environments, their implementation does not consider any kind of network emulation and has no mechanism for injecting faults into the system.

ELIoT~\cite{makinenELIoTDesignEmulated2017a} is based on Docker containers and supports the IoT protocols CoAP and LWM2M by using the open-source projects Leshan and coap-node.
While ELIoT includes the interaction with the environment for the use case described in the paper, this interaction is only modeled within the nodes (i.e., they implemented a simple calculation of an illuminance sensor value based on the time of day).
It does not integrate an environment emulation that would allow for two-way interaction between IoT systems and the environment. 

\subsubsection{Using Traffic Shaping.}

IOTier~\cite{Nikolaidis_IOTier_2021} is a virtual testbed for tiered IoT environments that is unfortunately not openly available. 
Nodes are represented via resource-constrained containers while networking is based on NetEm. %
A special focus is grouping emulated components into tiers with comparable capabilities, and enabling inter-tier as well as intra-tier communication.
It features a testbed controller in which operators can define desired runtime states over time. However, there is no API for integrating simulators of domain environments.
Its simulation engine uses fixed-increment time progression and can modify experiments via scheduled and conditional events.

Fogify~\cite{Symeonides_Fogify_2020} appears to be one of the most capable tools according to our criteria.
It uses Infrastructure-as-Code descriptions for containerized deployments to define experiment settings (i.e., Docker Compose) and features the possibility to adapt configurations at runtime (e.g., for injecting faults). %
Fogify uses Virtual eXtensible LAN (VXLAN) for overlay networks and is distributable across multiple physical hosts.
We classified this tool as being able to model energy consumption as this feature is described in the paper.
However, this is currently not implemented in code.
Although Fogify has an API for interacting with experiments during runtime, there is not yet a uniform way to integrate simulations of domain environments.

MockFog~\cite{hasenburgMockFogAutomatedExecution2021} is a tool for automated execution of fog application experiments. It consists of three modules: one for infrastructure setup, one for application management, and one for experiment orchestration, which enables the scripting of scenarios.
The experiment infrastructures are set up automatically in public cloud environments via dockerized application containers.
Hence, applications must support running inside Docker and must be available as container image.
Experiment descriptions can be used to generate events, such as traffic scenarios and network or machine failures. 

Blockade~\cite{blockade} is a test environment based on Docker containers and traffic shaping. The user creates a setup similar to a Docker Compose file and Blockade manages the set-up as well as tear-down processes. Each node is implemented as a separate Docker container. Blockade offers basic networking capabilities by using the Docker network and integrates the manipulation of network parameters via, e.g., NetEm settings.

Distem~\cite{sarzyniec2013design} is a virtual testbed using Linux Containers (LXC) that can be executed on multiple physical hosts.
One focus of Distem is resource allocation and assignment to achieve realistic setups for special devices (like IoT devices).
Network parameters can be adapted using NetEm.
Distem can be used via the command line and allows scriptable scenarios via its Ruby library.

\subsubsection{Using Software-Defined Networking.}

Fogbed~\cite{coutinhoFogbedRapidPrototypingEmulation2018}, as described by the original paper, uses Mininet for networking and is hence bound to a single host.
The latest prototype additionally extends MaxiNet, which enables emulating environments that span several physical machines.
Fogbed furthermore enables the testing of third-party systems such as resource management, virtualization, and service orchestration through standard interfaces.

EmuFog~\cite{mayerEmuFogExtensibleScalable2017a} is a fog computing emulation framework based on the distributable network emulator MaxiNet. The framework does not resort to simulations but is able to span an emulated network of thousands of virtual devices over multiple physical machines. EmuFog focuses on the networking components of fog computing by embedding a network topology generator, enhancer, and node placement algorithm. Applications have to be deployed as Docker containers.

\subsubsection{Using Network Simulation.}

Dockemu~\cite{petersenDockemuEvolutionNetwork2019} is the only tool without hardware integration that uses network simulation. It utilizes the network simulator ns-3 to model the communication between nodes, which in turn are represented by Docker containers. The paper recognizes the importance of providing realistic conditions and environmental factors for the applications under test. The tool itself, however, is restricted to controlling properties of nodes and the network but does not include mechanisms to provide a domain environment in which the application operates.

\subsection{Test Environments with Hardware Integration}
Next, we describe hybrid tools that offer the possibility to integrate real IoT devices in otherwise emulated environments to make experiments more realistic.

\subsubsection{Using Traffic Shaping.}

EmuEdge~\cite{zeng2019emuedge} is an openly available, hybrid simulator that can represent nodes using containers, virtual machines, and physical devices.
It supports OS-level as well as system-level virtualization and can interface simulators and real testbeds.
Networking is based on networking namespaces (\textit{netns}) and can replay real-world network traces.

Héctor~\cite{behnke2019hector} is an IoT testing framework with the main goal of representing devices realistically. Devices are emulated with QEMU in system mode, allowing fine grained performance moderation of individual devices and testing
on the target platform, including its corresponding microarchitecture. Specifically, this allows testing of devices that are not able to run Docker containers (e.g., microcontrollers). Physical as well as emulated devices can be part of the network, which itself can have emulated properties such as added delay and packet loss.

\subsubsection{Using Software-Defined Networking.}

Sendorek et al.~\cite{Sendorek_SDNTestbedForIoT_2018} describe an elaborated concept for a software-defined virtual test environment for IoT systems. Their system supports three so called "immersion levels" that range from fully virtualized environments for low-cost, scalable experiments to environments with real devices and sensors for testing under realistic conditions. The authors do not cover distributability or fault injection in their concept.

Chameleon~\cite{keaheyLessonsLearnedChameleon2020} builds upon OpenStack to deliver a testbed that can be used like a cloud. Chameleon is both a concept with an open-source implementation and a platform service supported by hardware at University of Chicago and at the Texas Advanced Computing Center that includes different nodes and setups including GPUs, FPGAs as well as ARM and x86 cores.
In addition to bare metal nodes, nodes virtualized with KVM can be used.
Besides the concept of an OpenStack-based testbed, the Chameleon project has some insights regarding the operational side of such a testbed, like user management, fair resource allocation with leases and lease reapers, security attacks etc.

StarBED~\cite{miyachi_starbed_2006} is a large-scale general purpose network testbed based on co-located physical nodes which uses SpringOS to build experiment topologies and drives experiments.
Its updated fourth version implements additional features, such as
wireless network emulation and a background traffic generator.
Although StarBED aims to enable Internet-scale experiments, it apparently lacks the possibility to emulate IoT characteristics (e.g., resource constraints, heterogeneous network capacities) and mainly acts as a resource management system. %

\subsubsection{Using Network Simulation.}

UiTiOt~\cite{ly-trongUiTiOtV3Hybrid2018a}, meanwhile in its third version, is a test environment for large-scale wireless IoT applications.
Instances of the application under test are executed using Docker Swarm on top of an OpenStack instance.
The network connections between the application instance (e.g., IEEE 802.11a/b/g, ZigBee) are emulated using the wireless emulator QOMET.
Apart from the virtual resources, UiTiOt can integrate physical nodes into the network.
The authors also introduce a web interface for users of the testbed and a load-balanced database for receiving and storing logs from the application under test.

WHYNET~\cite{zhouWHYNETHybridTestbed2006} is a hybrid testbed that focuses on mobile communication and applications, using a combination of simulation, emulation, as well as physical nodes and connections.
It simulates the network via the QualNet simulator and the sensor network simulation framework sQualNet, which is one of the few tools that model energy consumption.
Using the TWINE framework \cite{twine}, it emulates the network stack and the execution of applications to provide a scalable but realistic test environment. 
WHYNET includes a basic concept of mobility but does not allow the integration of domain-specific simulators for this purpose.

MobiNet~\cite{mahadevanMobiNetScalableEmulation2006} focuses on the evaluation of applications and network setup in ad hoc wireless networks. The tool allows the testing of different deployment schemes for applications and includes the simulation of movement of nodes. The core of MobiNet takes care of emulating the physical, data link, and network layers.
Edge nodes can be distributed across machines and can host multiple virtual nodes for large-scale environments. Unfortunately, the code for MobiNet is not publicly available.

\section{Discussion} %
\label{sec:discussion}

We identified themes that emerged in our comparison of test environments in each of our categories of characteristics: general characteristics, and those that relate to representation of nodes, network and domain environment.

\subsection{General} 
Testing of IoT systems is an active research area and many solutions try to help the developers of IoT systems in this regard. 
In our comparison, most systems were initially published within the last five years. 
These works include mature and widely adopted projects, but also ideas and research prototypes. 
Only two of the examined test environments are offered as a service.

\subsection{Nodes}
In our comparison we investigated the ability of test environments to use virtual and hardware nodes for the testing of IoT systems. 
For the virtual nodes, both containers and virtual machines are used, with recent works showing a tendency to use more lightweight container virtualization. 
This choice of virtualization type correlates with the integration of hardware nodes: 
Systems that include hardware nodes mostly use virtual machines, while the others mostly use containers. 
The ability to execute some nodes on actual hardware is missing from more than half of the test environments, even though this is especially important in many IoT use cases because often highly customized hardware is used.
While energy consumption modeling is crucial to test the behavior of battery-constrained IoT devices, this is barely considered in the tools covered. A better integration of power models, for example using simulators built for this purpose \cite{WiesnerThamsen_LEAF_2021}, would be an important next step for virtual test environments.

\subsection{Network}
The environments included in our comparison contain a mix of different methods to model the network. 
This includes two systems that do not even include the ability to specify a network topology, seven systems that support traffic shaping (usually via tc and NetEm), as well as full network simulation (four systems) and software defined networking (five systems). 
The scalability to large networks that need to be realized on multiple execution nodes is possible in almost all test environments that can distribute nodes. Network distributability only seems to still be a challenge when network simulators are used. 
Fault injection is an important feature for IoT testing, but dedicated support for defining and executing specific failure scenarios is missing from many IoT test environments.

\subsection{Domain Environment}
Despite the tight integration of distributed IoT systems with their environment, support for the simulation of domain environments is missing from all testing tools included in our comparison.
Accordingly, system developers have to resort to expensive and time-consuming field testing, when they want to test the interaction of IoT systems with their environment. %
While some environmental factors can be integrated in the testing by feeding applications recorded streams of sensor data, this integration is naturally limited and cannot model the manipulation of the environment by IoT systems.

\section{Research Outlook}
\label{sec:outlook}

While many tools exist that tackle the problem of testing distributed IoT systems, there are still important open challenges. 

\vspace{1mm}
\textbf{Research Gap.}
Currently, there is limited support for assessing key system requirements such as high resilience and low energy consumption.
However, the biggest gap in our view is the missing integration with domain environment simulations.
This integration is particularly important for IoT systems, because the tight coupling and interaction with the environment is a fundamental property of the Internet of Things.  
The integration of domain environment simulations like traffic or infrastructure simulations would allow for meaningful and continuous testing of these interactions.

\vspace{1mm}
\textbf{An Ideal IoT Test Environment.}
As we have derived the characteristics described in Section~\ref{sec:characteristics} from our understanding of the needs of a test environment for continuous testing, an ideal IoT test environment would fulfill all these characteristics.
Specifically, an ideal test environment would:
\begin{itemize}
\vspace{-.5em}
\item support testing on virtual and hardware nodes,
\item model and monitor the energy consumption,
\item include a network representation that allows complex network topologies and dynamic changes thereof, 
\item integrate domain environment simulations,
\item enable the distribution of nodes, networks, and domain environments across multiple physical nodes to allow the testing of large-scale deployments,
\item and also support fault injection on these three dimensions to evaluate the fault tolerance of the system under test.
\end{itemize}

\vspace{1mm}
\textbf{The Marvis Testing Framework.}
We are working on a framework towards our vision of an ideal IoT test environment, called Marvis~\cite{beilharzStagingEnvironmentInternet2021b}.
By combining virtual nodes (containers) with hardware nodes, Marvis offers capabilities for hybrid setups to combine the advantages of scalability and realism. 
Nodes can communicate via a simulated network realized by the network simulator ns-3. 

A focus of our work is the integration of domain environment simulators to enable the continuous testing of the often intricate interactions between the IoT system and the environment.
Currently, Marvis integrates the traffic simulator SUMO
to demonstrate this, allowing the testing of interactions between the real software systems that run on the nodes and the movement of road users in the traffic simulation. 
This integration is bidirectional, meaning both, the change of the movement of road users by the applications under test, and the change of connectivity in the network simulation by the traffic simulation is possible. 

Besides this, Marvis also offers fault injection in the three feature dimensions:
It is possible to inject faults in the nodes (e.g., start and stop nodes, or execute commands), the network simulation (e.g., connect or disconnect nodes, change network parameters like delay), or the domain-specific simulation (e.g., changing speed of vehicles).

\section{Related Work} %
\label{sec:rw}

Testing has been recognized as an important topic in IoT systems research since its beginning. Consequently, several related works provide an overview of testing research, environments and frameworks. 

Tonneau et al. \cite{tonneauHowChooseExperimentation2015} presented an extensive work focusing on the question of choosing the right wireless sensor network testing platform for specific environment characteristics in 2015.
It is the only related work in which all presented testbeds consist of devices carrying real sensors – no platforms were presented that only simulate or emulate devices under test.
Tonneau et~al. considered seven platform features: experimentation, scale, repeatability, mobility, virtualization, federation, and heterogeneity.

Dias et~al.~\cite{diasBriefOverviewExisting2018c} identified the motivation and challenges of testing planetary-scale, heterogeneous IoT applications and devices. Surveyed testing tools were chosen with no specific properties in mind, making 16 IoT testing platforms that were available in 2018 part of the survey. Tools were compared based on ten properties, including supported IoT layers, test level, test method, supported platforms, and scope (market/academic). The authors conclude that further research and development in the area of IoT testing is necessary, given the criticality of many IoT systems and the challenges of testing them.

A journal article from the same year by Chernyshev et~al.~\cite{chernyshevInternetThingsIoT2018} discusses the state of IoT research, simulators and testbeds. They defined a set of relevant research topics, including eight goals for the IoT. Furthermore, they performed a comparative study of nine simulation tools, categorized by the scope of coverage of the IoT architecture layers, as well as a comparison of three large-scale IoT hardware testbeds. They identified three open challenges concerning IoT testing: A lack of support for common IoT communication standards, a lack of end-to-end service simulation across all IoT layers, and a large discrepancy between simulator and real-world test results.

Patel et al.~\cite{patelSimulatorsEmulatorsTestbeds2019} compared a total of 26 simulators, emulators, and physical testbeds for the IoT. The authors discussed these groups of test environments independently from each other on characteristics such as scope, scale, and security measures. While there is no specific survey system or selection method given, the comparison is followed by a short analysis of the usage of simulators, emulators, and physical testbeds in the different stages of software development. 

Bures et al.~\cite{buresInteroperabilityIntegrationTesting2020} performed a systematic mapping study on interoperability and integration testing of IoT systems. Rather than comparing specific tools, they analyzed 115 out of 803 identified papers in the general area of IoT. The literature study was guided by seven research questions regarding research trends, researchers, publication media, topics, challenges, and limitations mentioned in the surveyed works. They conclude that there is a need for more specific testing methods for IoT systems.

\section{Summary}
\label{sec:summary}
This paper presented the current state of the art in continuous testing of distributed IoT systems.
Specifically, we described desirable characteristics for test environments in this context and compared IoT test environments that allow to run system code on virtual nodes.
Many solutions have been put forward, implementing various approaches to providing execution hosts and realizing network conditions. 
However, no currently available solution provides support for domain simulations, even though IoT systems form cyber-physical systems that make sense of and interact with their  surroundings.

We believe that systems that monitor and affect the real world should be tested comprehensively, especially in critical application domains such as traffic management, patient monitoring, and manufacturing.
Future work should therefore focus on providing comprehensive test environments, including simulation of domains and modeling of system characteristics such as energy consumption.

\bibliographystyle{splncs04}
\bibliography{bibliography}
\end{document}